%% file: arxiv_main.tex
\documentclass[cameraready]{Interspeech}
\title{Child-Centric Voice Anonymization in Single and Multi-Speaker Speech via Domain-Adapted SSL Models}

\author[affiliation={1}, orcid=0009-0009-5284-6714]{Pranav}{Tushar}
\author[affiliation={2}, orcid=0000-0002-6645-6524, correspondingauthor]{Xiao Xiao}{Miao}
\author[affiliation={1}, orcid=0000-0003-3410-8354]{Rong}{Tong}

\address{
    $^1$ Singapore Institute of Technology, Singapore \\
    $^2$ Duke Kunshan University, China 
}

\email{
tpranav2001@gmail.com,
pranav.tushar@singaporetech.edu.sg,
xiaoxiao.miao@dukekunshan.edu.cn,
tong.rong@singaporetech.edu.sg
}


\usepackage{booktabs}
\usepackage{multirow}
\usepackage{hyperref}
\usepackage{colortbl}
\usepackage{url}
\usepackage{makecell}

\begin{document}

\maketitle


\begin{abstract}
Voice anonymization aims to protect speaker identity while preserving linguistic content and speech usability. However, most anonymization systems are developed on adult speech, leading to degraded performance when applied to child speech. This paper investigates child-centric anonymization by adapting a self-supervised learning (SSL) based anonymization pipeline to the child speech domain. The system is adapted using child speech from the MyST corpus and evaluated under both single-speaker and two-speaker mixture conditions. Experimental results show that child-domain adaptation improves intelligibility and perceptual quality while maintaining strong privacy protection. Extending the approach to multi-speaker further demonstrates that combining target speaker extraction with child-adapted anonymization provides privacy protection while preserving conversational structure. These findings highlight the importance of child-specific adaptation for practical speech anonymization systems. 
\end{abstract}

\section{Introduction}

Voice anonymization aims to suppress speaker identity while preserving linguistic content and downstream usability. The VoicePrivacy Challenge~\cite{tomashenko2020introducing, panariello2024voiceprivacy,tomashenko2026third} has established common evaluation protocols and reference systems, accelerating progress in both signal-processing and neural approaches~\cite{patino2021speaker, panariello2024speaker, fang2019speaker, miao2022language}. However, most systems in this line of work are developed and evaluated on adult speech, and key components in modern pipelines, including content encoders and neural vocoders, are typically trained on adult data.

Children's speech constitutes a distinct acoustic domain, characterized by higher fundamental frequency, greater articulatory and prosodic variability, and developmental disfluencies. These properties create a substantial mismatch for adult-trained anonymization pipelines. Recent evaluations show that while privacy protection can remain comparable under standard automatic speaker verification (ASV) attackers, intelligibility and perceptual quality degrade substantially when adult-oriented systems are applied to child speech~\cite{kulkarni2025children}, motivating child-adapted anonymization.

Additionally, existing approaches suffer from two limitations. First, many systems anonymize children's speech by converting it toward adult-like voices, distorting age-dependent acoustic cues relevant to child-centered applications ~\cite{kulkarni2025children, dhar2026speaker}. Second, prior work considers only single-speaker scenarios, whereas real-world child speech, such as classroom interactions \cite{tushar2026personalized} or clinical sessions, frequently involves conversations between children and adults. In such multi-speaker settings, it is desirable to selectively anonymize the target child while optionally preserving or separately anonymizing the adult speaker.

These limitations motivate child-to-child anonymization, where speaker identity is obfuscated while preserving perceived child-specific acoustic characteristics, producing naturally childlike anonymized speech under both single- and multi-speaker 
conditions. Specifically, we adopt a self-supervised learning (SSL)-based anonymization system~\cite{miao2022language} 
that (i) decomposes the original 
speech into a content representation~\cite{van2022comparison}, a prosody representation using a pitch extractor, and a speaker 
representation~\cite{desplanques2020ecapa}; (ii) anonymizes the source speaker identity via a selective anonymization approach~\cite{select}, where the source speaker embedding is replaced by a reference embedding drawn from a speaker pool; 
and (iii) reconstructs anonymized speech by feeding the unchanged content and prosody representations together with the anonymized speaker embedding into a HiFi-GAN vocoder~\cite{kong2020hifi}. Since this system is originally trained on adult speech, we fine-tune both the content encoder and the vocoder on the MyST child speech corpus~\cite{pradhan2024my}, and replace the adult speaker pool with a controlled child speaker pool constructed from AI-generated voices screened for naturalness and age consistency, thereby improving utility on child speech and enabling child-to-child anonymization. We further extend this single-speaker system to multi-speaker conditions by combining target speaker extraction with the anonymization pipeline.

\begin{figure}
    \centering
\includegraphics[width=1\linewidth, trim=50 30 20 0, clip]{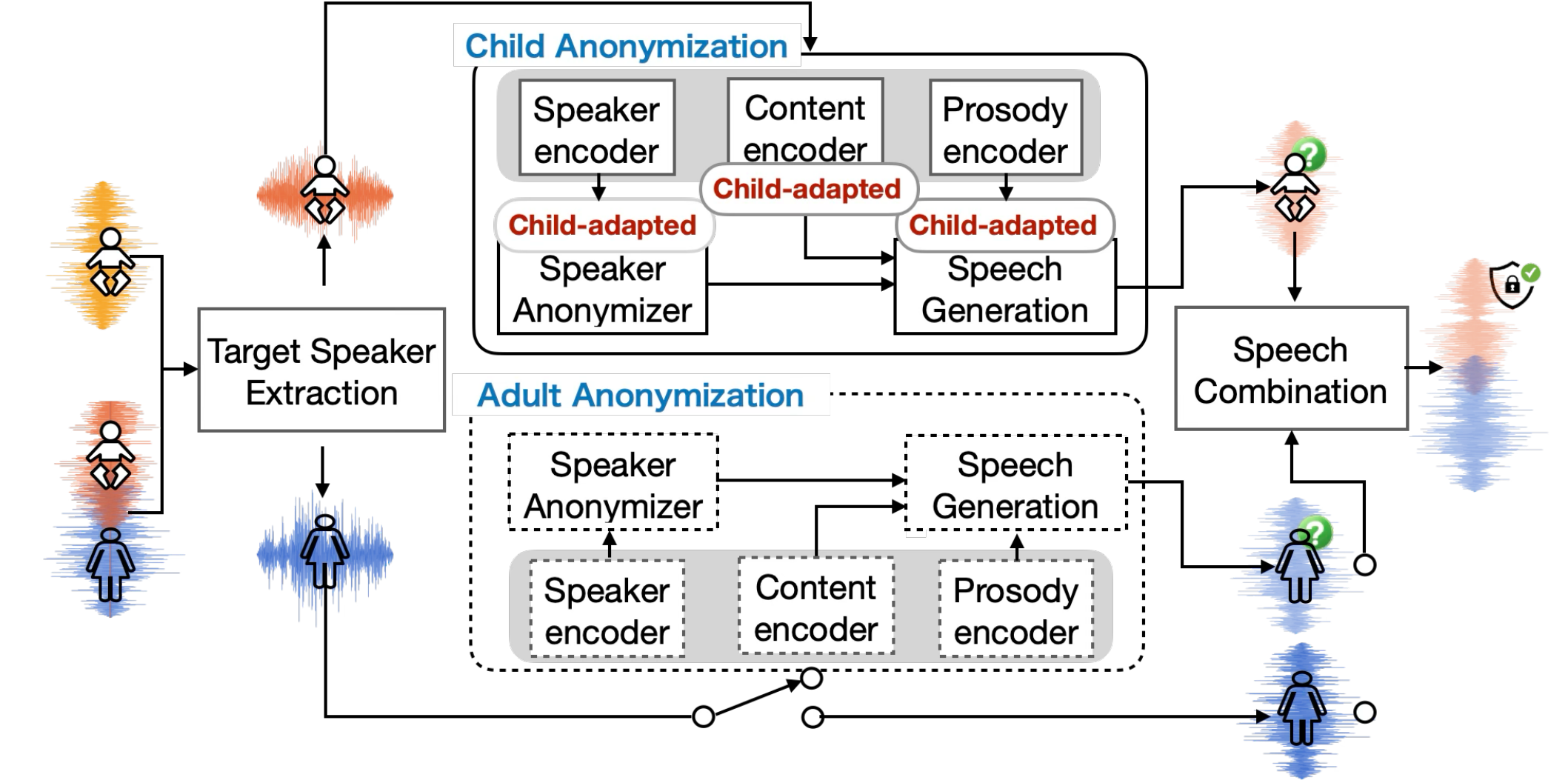}
    \caption{Child-centric anonymization.}
    \label{fig:system_paper}
\end{figure}

For single-speaker evaluation on both in-domain and zero-shot cross-accent benchmarks, the child-adapted system achieves strong privacy protection while maintaining competitive intelligibility and perceptual quality. For multi-speaker evaluation across three age-group pairings, adult--adult (AA), child--adult (CA), 
and child--child (CC), anonymization remains effective across all overlap levels, intelligibility degradation is moderate for AA mixtures but increases substantially for CA and CC conditions, reflecting the additional challenge of child target speaker 
extraction. An overview of the proposed pipeline is shown in Fig.~\ref{fig:system_paper}. 
The main contributions\footnote{Source code, pretrained models, and audio samples are publicly available at \url{https://github.com/pranavtushar/SSL-CVA}.} of this work are:


\begin{itemize}
    \item A systematic study of child-domain adaptation of SSL-based anonymization pipelines for child-to-child voice anonymization, evaluated on in-domain and zero-shot cross-accent benchmarks across privacy, intelligibility, perceptual 
    quality, and perceived speaker type.
    \item Extension to two-speaker mixtures, demonstrating that privacy remains relatively stable across overlap levels while intelligibility is primarily constrained by 
    target speaker extraction, particularly for child--child mixtures.
\end{itemize}

\section{Child-centric anonymization}
\subsection{SSL-based single-speaker child anonymization}
\label{sec:single-anon}
Given an input waveform $x$, we extract three disentangled representations: (i)~soft content features $c$ via a HuBERT-based encoder~\cite{van2022comparison}, (ii)~pitch contour $f_0$, and (iii)~a speaker embedding $s$ via ECAPA-TDNN~\cite{desplanques2020ecapa}. The content and prosody representations are kept intact to preserve linguistic information, while the speaker embedding undergoes an identity transformation to produce $\tilde{s}$. A HiFi-GAN vocoder~\cite{kong2020hifi} then resynthesizes the anonymized waveform $\tilde{x}$ from $(c, f_0, \tilde{s})$.
The baseline anonymization system is trained entirely on adult speech, leading to 
failure on child input at multiple stages. Prior work on child speech modeling 
reports consistent gains from domain adaptation of foundational speech 
models~\cite{fan2024benchmarking}. Motivated by this, we adapt the following 
components:

\noindent
\textit{HuBERT Content Encoder.} The HuBERT soft content encoder, pretrained on 
adult speech, yields suboptimal linguistic representations for child utterances due to the acoustic mismatch in pitch range, formant frequencies, and speaking rate. We therefore fine-tune it on the MyST child speech corpus~\cite{pradhan2024my} to improve the extraction of linguistically meaningful representations from child speech.

\noindent
\textit{Child Speaker Pool.} In the original system, the source speaker embedding 
$s$ is replaced by a reference embedding $s_{\mathrm{ref}}$ drawn from an adult 
speaker pool, i.e.\ $\tilde{s} = s_{\mathrm{ref}}$. For child-to-child 
anonymization, we replace this pool with a controlled set of child-like voices, 
where $s_{\mathrm{ref}}$ is sampled to preserve age-consistent acoustic 
characteristics. The child speaker pool is constructed from AI-generated voices 
screened for naturalness and age consistency.
Note that the ECAPA-TDNN speaker encoder is kept fixed, as the anonymized identity is obtained by directly replacing the source embedding with a reference embedding from the child speaker pool, which primarily determines the anonymized speaker identity.

\noindent
\textit{HiFi-GAN Vocoder.} The HiFi-GAN vocoder, also trained on adult waveforms, 
tends to shift resynthesized child speech toward adult-like acoustics, distorting 
age-dependent cues such as pitch range and vocal tract resonances. We fine-tune it on the same MyST corpus to enable faithful resynthesis that preserves child-specific spectral and prosodic characteristics.


\subsection{Target child anonymization in multi-speaker speech}
We extend child-centric anonymization to two-speaker mixtures. Given a mixture $x_{\mathrm{mix}}$ containing a target speaker and a non-target speaker, along with a short reference utterance $r_{\mathrm{target}}$, the goal is to anonymize only the target speaker's identity while leaving the non-target speech unmodified.

The pipeline operates in three stages. First, a Conformer-based target speaker extraction (TSE) model~\cite{liu2024target, liu2024libri2vox} isolates the target signal. The model estimates the target speaker's complex spectrum from the STFT of $x_{\mathrm{mix}}$, conditioned on a speaker embedding derived from $r_{\mathrm{target}}$, and inverts it to obtain $\hat{x}_{\mathrm{target}}$. The non-target signal is recovered as the residual $\hat{x}_{\mathrm{non\text{-}target}} = x_{\mathrm{mix}} - \hat{x}_{\mathrm{target}}$. Second, the extracted target signal $\hat{x}_{\mathrm{target}}$ is anonymized using the single-speaker pipeline described in Section \ref{sec:single-anon} with the child-adapted configuration for child targets and the base configuration for adult targets, producing $\tilde{x}_{\mathrm{target}}$. Third, the anonymized mixture is reconstructed by replacing the target component $
    \tilde{x}_{\mathrm{mix}} = \tilde{x}_{\mathrm{target}} + \hat{x}_{\mathrm{non\text{-}target}}$.

We evaluate across three age pairings: adult--adult (AA), child--adult (CA), and child--child (CC) to reflect diverse real-world scenarios. This distinction is important because both extraction and anonymization become more challenging when speakers are acoustically similar, as in CC pairs.

\section{Experiments}
\subsection{Datasets and evaluation metrics}

\noindent
\textit{Single-Speaker Datasets:}
Table~\ref{tab:datasets} summarizes the datasets used for adaptation and evaluation. MyST~\cite{pradhan2024my} serves as the in-domain child speech corpus; we use the training partition for child-domain adaptation and reserve the test partition for evaluation. Zero-shot generalization is assessed on cross-accent English datasets (MPS~\cite{gothi2024dataset}, SpeechOcean~\cite{zhang2021speechocean762}). 

\begin{table}[t]
\centering
\caption{Datasets used for single speaker experiments.}
\label{tab:datasets}
\resizebox{\columnwidth}{!}{%
\begin{tabular}{lllr}
\toprule
\textbf{Dataset} & \textbf{Language} & \textbf{Domain} & \textbf{Use} \\
\midrule

MyST~\cite{pradhan2024my} & English (US) & In-domain & \makecell[r]{Train+Dev: adapt \\ Test: eval} \\
MPS~\cite{gothi2024dataset} & English (Indian acc. ) & Cross-accent & Zero-shot eval \\
SpeechOcean~\cite{zhang2021speechocean762} & English (Mandarin acc.) & Cross-accent & Zero-shot eval \\

\bottomrule
\end{tabular}%
}
\end{table}

\begin{table}[t]
\centering
\caption{Evaluation metrics and protocols for single and multi-speaker experiments.}
\label{tab:metrics}
\resizebox{\columnwidth}{!}{%
\begin{tabular}{lll}
\toprule
\textbf{Metric} & \textbf{Backend} & \textbf{Protocol} \\
\midrule

\multicolumn{3}{c}{\textbf{Single-speaker evaluation}} \\
\midrule
EER ($\uparrow$) & ECAPA-TDNN ASV\footnote{\url{https://huggingface.co/speechbrain/spkrec-ecapa-voxceleb}}~\cite{chung2018voxceleb2} & Speaker verification (original vs.\ anonymized) \\
WER ($\downarrow$) & Whisper Large-v3\footnote{\url{https://huggingface.co/openai/whisper-large-v3}}~\cite{radford2023robust} & ASR error on anonymized speech \\
NISQA-MOS ($\uparrow$) & NISQA\footnote{\url{https://github.com/gabrielmittag/NISQA}}~\cite{mittag2021nisqa} & Objective speech quality \\
Human eval. & Listening test & Naturalness, fluency, similarity, age \\

\midrule
\multicolumn{3}{c}{\textbf{Multi-speaker evaluation}} \\
\midrule
EER ($\uparrow$) & ECAPA-TDNN ASV~\cite{chung2018voxceleb2} & Target verification (OO: orig–orig, OA: orig–anon) \\
tWER ($\downarrow$) & \texttt{gpt-4o-transcribe-diarize} & Target WER on diarized target channel \\
DER ($\downarrow$) & DiariZen + pyannote~\cite{han2025fine,Bredin23} & Diarization error vs.\ oracle RTTM \\

\bottomrule
\end{tabular}%
}
\end{table}

\noindent
\textit{Multi-Speaker Mixtures:}
We construct two-speaker mixtures following a SparseLibriMix-style procedure~\cite{cosentino2020librimix}. Child speech is drawn from the MyST test split and adult speech from LibriSpeech test-clean. Utterances are cropped or padded to 5\,s and mixed at 0\,dB SNR with overlap ratios of 0--100\% in 20\% increments. The first speaker is designated as the anonymization target. We generate 50 mixtures per overlap level per age-group pairing (AA,CA,CC), yielding 900 mixtures in total.

\noindent
\textit{Evaluation Metrics:}
Table~\ref{tab:metrics} summarizes the evaluation metrics and protocols used in this work. 
For single-speaker experiments, we additionally conduct a listening study with 13 participants who rate naturalness, fluency, speaker similarity, and perceived age. Six source utterances are sampled across the evaluation datasets (two per dataset) to cover both short and longer child speech segments. For each source utterance, listeners hear the original signal and five anonymized variants and rate naturalness, fluency, speaker similarity, and perceived age. Each participant evaluates 30 anonymized samples in total.

For multi-speaker experiments, pseudo-reference transcripts are generated using \texttt{gpt-4o-transcribe-diarize}\footnote{\url{https://developers.openai.com/api/docs/models/gpt-4o-transcribe-diarize}} to obtain speaker-attributed hypotheses for the target channel. Pseudo-references are used because ground-truth transcripts are not available after mixture construction, target extraction, and reconstruction. While this introduces a potential source of ASR error and may affect absolute WER values, it provides a consistent proxy for comparing relative intelligibility trends across systems and overlap conditions.

\subsection{Experimental setup}
For child-domain adaptation, we fine-tune components of the \textbf{SSL}-based anonymization pipeline, namely the HuBERT content encoder and the HiFi-GAN vocoder, on the MyST corpus following the strategy in~\cite{miao2022language}. 
We evaluate four configurations to analyze the effect of adaptation: Base/Base, FT/Base, Base/FT, and FT/FT. The reference speaker pool is constructed from AI-generated child-like voices using Typecast\footnote{\url{https://typecast.ai/}} and SpeechGen\footnote{\url{https://www.speechgen.app/}}, filtered for naturalness and age consistency. The pool contains 44 utterances from 16 synthetic child-like speakers. Samples are manually screened to ensure naturalness and consistent child-like vocal characteristics before embedding extraction. In addition to the \textbf{SSL}-based anonymization approach, we also consider the signal-processing baseline \textbf{B2}~\cite{patino2021speaker}, which applies the McAdams coefficient and has been shown to better preserve child-specific characteristics~\cite{dhar2026speaker}. The multi-speaker pipeline chains target speaker extraction, single-speaker anonymization, and mixture reconstruction. We employ a Conformer-based TSE model~\cite{liu2024target, liu2024libri2vox} trained on adult speech without child adaptation, allowing us to analyze domain mismatch effects under CA and CC conditions. For child targets, we apply the child-adapted (FT/FT) anonymization configuration; for adult targets, the base configuration.

\begin{table}[t]
  \caption{MyST (in-domain) component study for the selective anonymization system.}
  \label{tab:in_domain}
  \centering
  \footnotesize
  \setlength{\tabcolsep}{4pt}
  \renewcommand{\arraystretch}{1.05}
  \begin{tabular}{llccc}
    \toprule
    \textbf{Soft Encoder} & \textbf{HiFi-GAN} &
    \textbf{EER} ($\uparrow$) &
    \textbf{WER} ($\downarrow$) &
    \textbf{NISQA-MOS} ($\uparrow$) \\

        \midrule
        \rowcolor{gray!15}
        Base & Base & 43.80 & 17.31 & 3.60 \\
        FT   & Base & 38.10 & 19.53 & \textbf{3.70} \\
        Base & FT   & 40.68 & 20.67 & 3.10 \\
        \rowcolor{gray!15}
        FT   & FT   & \textbf{45.09} & \textbf{16.64} & 3.36 \\
        \bottomrule

  \end{tabular}
\end{table}

\begin{table*}[t]
\caption{Privacy (EER), intelligibility (WER), and perceptual quality (NISQA-MOS) across child speech datasets. Best results per dataset and metric are shown in bold. }
\label{tab:eer_wer_mos_child}
\centering
\footnotesize
\begin{tabular}{l l | c c c c | c c c c | c c c c}
\toprule
\multirow{2}{*}{\textbf{Dataset}} &
\multirow{2}{*}{\textbf{Age}} &
\multicolumn{4}{c|}{\textbf{EER (OA) \% $\uparrow$}} &
\multicolumn{4}{c|}{\textbf{WER \% $\downarrow$}} &
\multicolumn{4}{c}{\textbf{NISQA-MOS $\uparrow$}} \\
\cmidrule(lr){3-6} \cmidrule(lr){7-10} \cmidrule(lr){11-14}
& &
\textbf{Org} & \textbf{B2} & \textbf{SSL-B} & \textbf{SSL-FT} &
\textbf{Org} & \textbf{B2} & \textbf{SSL-B} & \textbf{SSL-FT} &
\textbf{Org} & \textbf{B2} & \textbf{SSL-B} & \textbf{SSL-FT} \\
\midrule
\textbf{MyST (test)} & age (8-11) &
15.39 & 42.10 & 43.80 & \textbf{45.09} &
14.55 & 20.02 & 17.31 & \textbf{16.64} &
2.65 & 2.25 & \textbf{3.60} & 3.36 \\
\midrule
\multirow{2}{*}{\textbf{SpeechOcean}} & age (6-10) &
4.32 & 35.71 & 34.97 & \textbf{39.88} &
27.26 & 55.57 & \textbf{41.99} & 43.36 &
3.37 & 2.29 & \textbf{3.31} & 2.85 \\
& age (11-15) &
2.24 & 35.42 & 37.74 & \textbf{38.46} &
12.03 & 36.14 & \textbf{21.38} & 23.61 &
3.40 & 2.31 & \textbf{3.38} & 2.97 \\
\midrule
\textbf{MPS} & age (7-11) &
0.01 & 31.44 & 39.50 & \textbf{40.94} &
12.68 & 18.31 & 16.20 & \textbf{15.72} &
2.15 & 1.75 & \textbf{3.31} & 2.65 \\
\bottomrule
\end{tabular}
\end{table*}

\begin{figure}
    \centering
    \includegraphics[width=0.75 \linewidth]{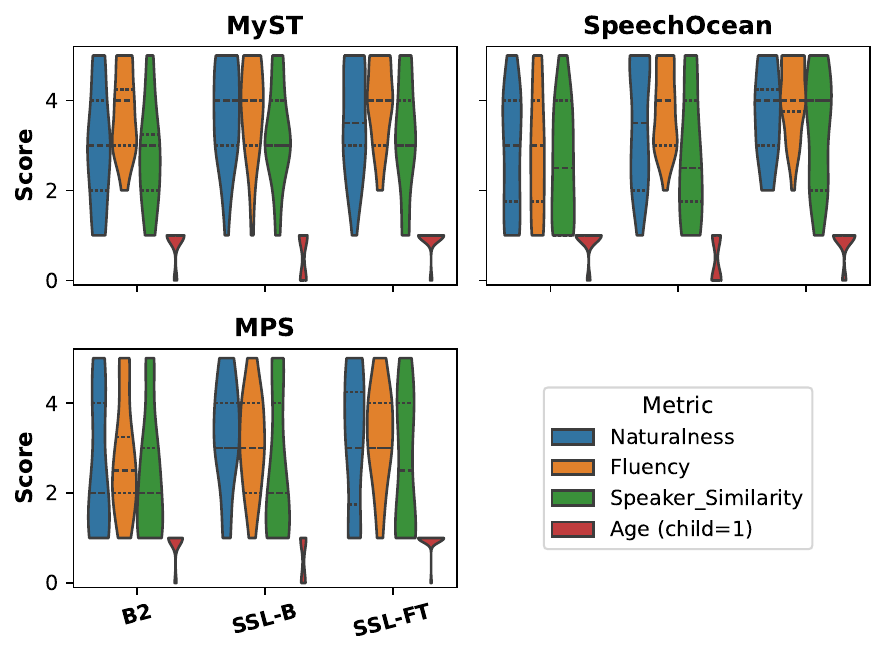}
    \caption{Subjective evaluation across datasets: listener ratings for naturalness, fluency, speaker similarity  \& perceived childness.}
    \label{fig:human_eval_violin}
\end{figure}

\subsection{Results and discussion}

\subsubsection{Single-speaker anonymization.}

\noindent
\textit{In-domain (MyST).} Table~\ref{tab:in_domain} shows that adapting only one component (FT/Base or Base/FT) degrades performance, likely due to representation-synthesis mismatch. In contrast, 
full child-domain adaptation (FT/FT) achieves the best trade-off, improving EER 
from 43.80\% to 45.09\% and reducing WER from 17.31\% to 16.64\%, although its NISQA-MOS is slightly lower than Base/Base. Therefore, we adopt the fully adapted (FT/FT) configuration for all subsequent evaluations, denoted as SSL-FT.

\noindent
\textit{Cross-domain generalization.}
Table~\ref{tab:eer_wer_mos_child} shows that SSL-FT achieves the strongest privacy protection, obtaining the highest EER across all datasets. In terms of intelligibility, SSL-FT yields the lowest WER on MyST and MPS, and remains competitive on SpeechOcean, demonstrating that child-domain adaptation effectively preserves speech utility beyond the training domain. Although SSL-FT does not consistently improve NISQA-MOS relative to SSL-B, it achieves stronger privacy protection and competitive intelligibility across datasets. 


\noindent
\textit{Human evaluation.}
Fig.~\ref{fig:human_eval_violin} summarizes subjective ratings across naturalness, fluency, and speaker similarity on a 1--5 Likert scale, where higher scores indicate better naturalness/fluency and greater perceived speaker similarity. For age perception, listeners performed a binary judgment (child vs.\ adult), and the reported value corresponds to the proportion of samples perceived as child speech (child = 1). The SSL-based anonymization systems (SSL-B and SSL-FT) receive higher naturalness and fluency ratings than the B2 baseline, indicating improved perceptual quality of the synthesized speech. Listener judgments further show that the child-adapted configuration (SSL-FT) more consistently preserves perceived childness while maintaining low speaker similarity scores. These results are consistent with the objective results, confirming that child-adapted neural anonymization can achieve effective identity concealment without substantially altering age-related acoustic characteristics. 

\subsubsection{Target children anonymization in multi-speaker speech}
\label{sec:multispeaker_results}

\begin{figure}
    \centering
    \includegraphics[width=1\linewidth, trim=10 10 30 0, clip]{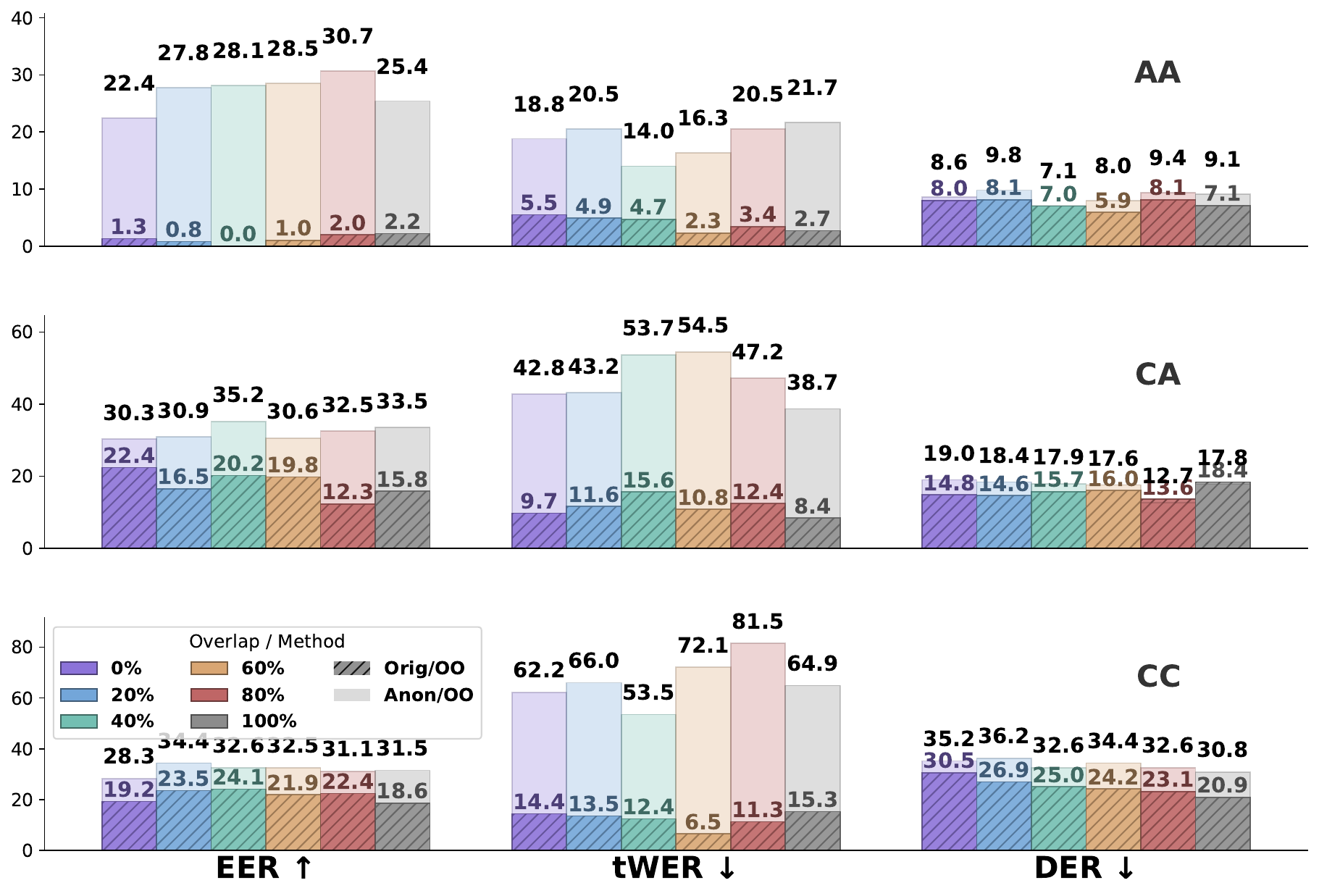}
    \caption{Multi-speaker evaluation: tWER, DER, and EER (\%) by age-group pairing (AA/CA/CC) and overlap ratio. OO denotes original--original verification trials and OA denotes original--anonymized verification trials.}
    \label{fig:combined_multispeaker}
    \vspace{-7mm}
\end{figure}

Fig.~\ref{fig:combined_multispeaker} summarizes target-speaker privacy (EER), intelligibility (tWER), and conversational-structure preservation (DER) across overlap ratios and age-group pairings.

\noindent\textit{Privacy is robust to overlap.}
The first cluster of Fig.~\ref{fig:combined_multispeaker} shows the EER under different overlap ratios. After SSL-based anonymization, OA EERs (solid bars) are consistently and comfortably higher than OO EERs (hatched bars) across all age-group pairings, with EER increasing relative to the OO condition across all overlap ratios and subsets (AA/CA/CC), indicating effective identity suppression even when the target speaker is first extracted from a mixture. Importantly, OA EER remains within a relatively narrow range as the overlap ratio increases from 0--100\%, suggesting that anonymization effectiveness remains relatively stable across overlap levels. However, because anonymization operates on extracted target speech, privacy outcomes remain coupled to extraction quality. 

\noindent\textit{Utility degrades with separation difficulty, especially for child--child mixtures.}
The second cluster of Fig.~\ref{fig:combined_multispeaker} shows that tWER increases with overlap and shows a strong dependence on age-group pairing. AA mixtures remain the easiest regime (low baseline tWER and moderate increase after anonymization), while CA exhibits intermediate degradation. CC mixtures are consistently the most challenging: even at low overlap, tWER is higher than AA and CA, and it increases sharply under heavier overlap. DER follows a similar trend, with the largest diarization errors observed in CC conditions, reflecting the difficulty of separating acoustically similar child voices. The gap between CA and CC further emphasizes that child-specific acoustics (higher pitch and rapid spectral dynamics) amplify extraction ambiguity, making CC the critical stress-test condition.


Overall, the results highlight a privacy--utility decoupling in multi-speaker child scenarios: privacy metrics remain relatively stable across overlap levels, while downstream utility is primarily bounded by extraction quality, particularly for child--child mixtures. This suggests that improving child-robust target speaker extraction is likely the most direct path to improving conversational child anonymization without weakening privacy.

\section{Challenges and limitations}

Children’s speech differs substantially from adult speech, introducing challenges for both modeling and evaluation. Child recordings often contain higher background noise, classroom reverberation, microphone variability, and spontaneous vocal behaviors such as disfluencies and non-lexical vocalizations. These factors can create mismatches between transcripts and spoken content. As a result, ASR-based metrics such as WER may partially reflect transcription variability rather than purely anonymization-induced degradation.

Another limitation concerns the adult-centric nature of several evaluation models used in the pipeline. Although the anonymization system is partially adapted to child speech, multiple evaluation components remain primarily trained on adult data. These include ASR systems, diarization models, speaker verification attackers, and perceptual quality predictors. Consequently, some reported performance differences may reflect domain mismatch in the evaluation models rather than limitations of the anonymization framework itself. For example, MOS predictors such as NISQA are largely trained on adult speech, which may reduce reliability when evaluating child voices. Age preservation is also difficult to quantify automatically. The age prediction models we tested generalized poorly on our child datasets, so we relied on human perceptual judgments to assess perceived age consistency; this improves validity but limits scalability.



Finally, several components in the multi-speaker pipeline remain adult-trained, including the target speaker extraction model and the ASV attacker. This mismatch likely contributes to performance degradation in child–child mixtures and under high overlap conditions. Stronger attacker models could further probe anonymization robustness. Overall, these factors highlight the need for child-adapted evaluation models and stronger attacker settings to better characterize privacy–utility trade-offs in child speech anonymization.

\section{Conclusions and future work}

This work presented a child-centric study of voice anonymization through domain adaptation of self-supervised speech models. Adapting the HuBERT content encoder and HiFi-GAN vocoder to child speech improves intelligibility while maintaining strong privacy protection across datasets. Human evaluations further indicate that the proposed approach better preserves perceived childness compared to signal-processing baselines.
Extending the analysis to multi-speaker mixtures reveals a clear decoupling between privacy and intelligibility. Privacy remains relatively stable across overlap conditions, while intelligibility degradation is primarily driven by target speaker extraction errors, particularly in child–child mixtures.

These findings highlight the importance of child-aware anonymization strategies as speech technologies are increasingly deployed in child-centered applications. Because children’s speech contains sensitive biometric information, privacy-preserving processing should be carefully considered when developing speech technologies involving minors. Future work will explore child-adapted extraction and verification models, improved evaluation tools for child speech, and multilingual child speech anonymization using broader child speech corpora.

\section{Acknowledgments}
This work is supported by the Singapore Ministry of Education
(MOE) Academic Research Fund (AcRF) Tier 1 grant R-R13-
A405-0005.

\section{Generative AI Use Disclosure}

Generative AI tools were used in a limited manner for (i) language editing during manuscript preparation and (ii) transcription support for WER-based evaluation using OpenAI ASR models. In addition, AI-generated text-to-speech (TTS) voices (Typecast and SpeechGen) were used to construct the child-like reference speaker pool for selective anonymization; these samples were used only for reference embedding extraction and were not used to train the anonymization models. No generative AI system was used for experimental design, model development, or interpretation of results.

\bibliographystyle{IEEEtran}
\bibliography{mybib}

\clearpage
\appendix
\clearpage
\appendix
\input{appendix}

\end{document}

%% file: appendix.tex
\section{Additional Experimental Details}

\subsection{Comparison of WER}

The multi-speaker experiments in the main paper evaluate target-speaker intelligibility using pseudo-reference transcripts generated by \texttt{gpt-4o-transcribe-diarize}. To verify that the reported conclusions are not dependent on the choice of reference transcripts, we additionally compute WER using transcript-grounded references derived from Montreal Forced Aligner (MFA) \cite{mcauliffe2017montreal}.

\subsubsection{Protocol}
We generated 900 deterministic two-speaker mixtures across adult--adult (AA), child--adult (CA), and child--child (CC) conditions at overlap ratios of 0\%, 20\%, 40\%, 60\%, 80\%, and 100\%. Whisper Large-v3 was used to transcribe the anonymized target channel (\texttt{target\_anon.wav}) for both evaluation protocols. The only difference lies in the reference transcript: (i) \textbf{Pseudo Ref.}, which uses pseudo-reference transcripts generated by \texttt{gpt-4o-transcribe-diarize} (as reported in the main paper), and (ii) \textbf{GT Ref.}, which uses transcript-grounded references derived from the original source utterances using MFA. Prior to WER computation, transcripts were normalized by removing punctuation and annotation artifacts, expanding common contractions, and standardizing numeric expressions where applicable. Because Whisper occasionally produced long hallucinated hypotheses for short child-speech segments, per-utterance WER was capped at 100\% before averaging within each subset and overlap condition.

\subsubsection{Discussion}
Table~\ref{tab:wer_compare} shows that although the absolute WER values differ between the pseudo-reference and MFA transcript-grounded evaluation protocols, both exhibit the same qualitative trends. Across all overlap conditions, AA consistently achieves the lowest WER, whereas CA and CC remain substantially more challenging. In particular, CC yields the highest WER across most overlap settings, indicating that target speaker extraction becomes increasingly difficult when both speakers share similar child-specific acoustic characteristics. The consistency between the two evaluation protocols provides additional evidence that the conclusions reported in the main paper are not driven by the pseudo-reference generation procedure.

\vspace{-0.7em}
\begin{table}[!ht]

\footnotesize
\centering
\caption{Comparison of target-speaker WER (\%) computed using pseudo-reference (Pseudo Ref.) and MFA transcript-grounded (GT Ref.) reference transcripts. The pseudo-reference WER values are reproduced from Fig.~3 of the main paper for ease of comparison. Although the absolute values differ, both evaluation protocols exhibit similar qualitative trends across AA, CA, and CC conditions. Lower values indicate better intelligibility ($\downarrow$).}
\label{tab:wer_compare}

\begin{tabular}{c|cc|cc|cc}
\toprule
&
\multicolumn{2}{c|}{AA} &
\multicolumn{2}{c|}{CA} &
\multicolumn{2}{c}{CC} \\
\textbf{Overlap (\%)} &
\textbf{Pseudo Ref.} &
\textbf{GT Ref.} &
\textbf{Pseudo Ref.} &
\textbf{GT Ref.} &
\textbf{Pseudo Ref.} &
\textbf{GT Ref.} \\
\midrule
0   & 18.8 & 20.85 & 42.8 & 41.35 & 62.2 & 59.15 \\
20  & 20.5 & 20.59 & 43.2 & 48.76 & 66.0 & 64.28 \\
40  & 14.0 & 17.76 & 53.7 & 49.26 & 53.5 & 59.91 \\
60  & 16.3 & 20.09 & 54.5 & 40.98 & 72.1 & 69.35 \\
80  & 20.5 & 22.74 & 47.2 & 43.42 & 81.5 & 71.38 \\
100 & 21.7 & 23.61 & 38.7 & 41.57 & 64.9 & 66.54 \\
\bottomrule
\end{tabular}
\normalsize
\end{table}